\documentclass[12pt,a4paper]{article}
\pdfoutput=1
\usepackage{a4wide}
\usepackage[centertags]{amsmath}
\usepackage{amssymb,amsthm}
\usepackage{tensor}
\usepackage{csquotes}
\usepackage{setspace}
\usepackage[british]{babel}
\usepackage{float}
\usepackage{bbold}
\usepackage[pdftex]{graphicx,color} 
\allowdisplaybreaks
\numberwithin{equation}{section}
\begin{document}

\thispagestyle{empty}

\begin{center}
{\Large  \bf Quantum Gravitational Corrections to the
Entropy of a Reissner-Nordstr\"om Black Hole } 
\end{center}

\vspace*{1 cm}

\centerline{Ruben Campos Delgado\footnote{email: ruben.camposdelgado@gmail.com}}
\vspace{1 cm}

\begin{center}{
Bethe Center for Theoretical Physics\\
{\footnotesize and}\\
Physikalisches Institut der Universit\"at Bonn,\\
Nussallee 12, 53115 Bonn, Germany}
\end{center}

\vspace*{1cm}

\centerline{\bf Abstract}
\vskip .3cm
Starting from an effective action for quantum gravity, we calculate the quantum gravitational corrections to the Wald entropy of a four dimensional non-extremal Reissner-Nordstr\"om (RN) black hole in the limit of small electric charge, generalising a previous calculation carried out by Calmet and Kuipers \cite{Calmet:2021lny} for a Schwarzschild black hole. We show that, at second order in the Ricci curvature, the RN metric receives quantum corrections which shift the classical position of the event horizon. We apply the Wald entropy formula by integrating over the perimeter of the quantum corrected event horizon. 
We then compute the quantum gravitational corrections to the temperature and the pressure of the black hole. 
\newpage
\section{Introduction}
One of the greatest achievement in theoretical physics in the last fifty years is the discovery that black holes have an entropy \cite{Hawking:1975vcx, Bekenstein:1973ur}. The leading classical contribution to the entropy of a black hole in Einstein’s general
relativity is equal to one quarter of the area of the event horizon. If the theory of gravity is modified, then the entropy receives additional contributions. Recently \cite{Calmet:2021lny}, effective field theory methods were used to calculate the quantum gravitational corrections to the entropy of a Schwarzschild black hole using the Wald entropy formula \cite{Wald:1993nt}, which in four dimensions reads
\begin{equation}\label{eq:wald}
    S_{\text{Wald}}=-2\pi \int\limits_{r=r_h} d\Sigma\, \epsilon_{\mu\nu}\epsilon_{\rho\sigma}\frac{\partial \mathcal{L}}{\partial R_{\mu\nu\rho\sigma}},
\end{equation}
where $\mathcal{L}$ is the Lagrangian density of the theory, $d\Sigma=r^2\sin\theta d\theta d\phi$, $R_{\mu\nu\rho\sigma}$ is the Riemann tensor and $\epsilon_{\mu\nu}$ is an antisymmetric tensor normalised as $\epsilon_{\mu\nu}\epsilon^{\mu\nu}=-2$. 
The integral has to be evaluated at the horizon radius $r_{h}$. An important observation is that the  metric itself may receive quantum gravitational corrections that could impact the position of the horizon. The Schwarzschild metric does not receive corrections at second order in curvature \cite{Calmet:2018elv}, but it does at third order \cite{Calmet:2021lny}. 

In this work we apply the same effective field theory approach to quantum gravity as in \cite{Calmet:2021lny} to compute the quantum gravitational corrections at second order in curvature to the entropy of a more realistic type of black holes, namely the electrically charged ones, also known as Reissner-Nordstr\"om (RN) black holes. The motivation behind adding electric charge is dictated not only by theoretical interest. Indeed, real astrophysical black holes do have a tiny electric charge, which can be approximately estimated as 100 Coulombs per solar mass \cite{Bally:1978ApJ}. If one is interested only in the motion of photons and neutral matter (neutrinos and gravitational waves) in the vicinity of the hole, then the assumption of zero charge is justified. However, the dynamics of charged particles like electrons and protons can be profoundly affected. The electric field of the black hole can do work and accelerate the particles to large, relativistic speeds. Incidentally, this could explain why cosmic rays are detected on the Earth with very high energies (see M. Zajacek, A. Tursunov, Electric charge of black holes: Is it really always negligible?, \texttt{arXiv:1904.04654 [astro-ph.GA]}). 

From the technical point of view, there is a difference between the result of \cite{Calmet:2021lny} and ours. Unlike the Schwarzschild black hole, which is a vacuum solution of general relativity, the presence of electric charge for the RN black hole implies a non-vanishing energy-momentum tensor. As a consequence, we found that the classical RN metric receives quantum gravitational corrections already at second order in curvature.  We obtained these corrections by solving the quantum corrected Einstein and Maxwell equations perturbatively in the gravitational coupling $G_N$ up to order $\mathcal{O}(G^2_N)$.  We present their derivation in Section \ref{sec:metric}.  
Quantum corrections to the RN metric were already considered in the literature using alternative methods. In \cite{Donoghue:2001qc, Holstein:2006ud} it was assumed that it is quantum matter, and not quantum gravity, which is responsible for the quantum effects. Accordingly, the energy-momentum tensor was expanded in a power series in the fine structure constant $\alpha$ using usual Feynman diagram techniques, then a Fourier transformation of the form factors was performed to get the correction to the metric. In \cite{Buric:1999wf}, quantum corrections to the RN metric were obtained within the context of 2D spherically symmetric dilaton gravity models.  

To compute the entropy, we restricted to the case of a non-extremal RN black hole and we additionally made the well-founded physical assumption that its charge $Q$ is very small compared to its mass $M$. To be precise, in every expressions we kept only terms up to order $\mathcal{O}\left(Q^2\right)$. In this sense, our results can be thought of a small correction to those obtained for the Schwarzschild black hole. As already mentioned, the Wald entropy requires to integrate over the event horizon.  We found that the corrections to the metric imply a small shift of the horizon radius, which contributes to the final expression of the entropy. These calculations are presented in Section \ref{sec:entropy}.
Previous works obtained the quantum corrections to the entropy of non-extremal RN black holes in other ways. Most notably, by computing the stringy $\alpha'$ corrections \cite{Cano:2019ycn}, by applying the brick-wall method \cite{Li:2000hd}, by counting the degrees of freedom near the horizon \cite{Yoon:2007aj}, or by immersing the black hole in an isothermal bath within a cavity \cite{Akbar:2003mv}.

After computing the entropy, in Section \ref{sec:firstlaw} we have a look at the first law of thermodynamics. We found that both the temperature and the pressure of RN black holes receive quantum gravitational corrections at order $\mathcal{O}\left(Q^2\right)$ in such a way that the first law remains valid.  
\section{Quantum gravitational corrections to the Reissner-Nordstr\"om metric}\label{sec:metric}
\subsection{Equations of motion from the effective action}\label{sec:equations_metric}
Effective field theory is an organised procedure that separates out the effects of high energy from those at low energy. Any ultra-violet completion of quantum gravity that reduces to general relativity in its low energy regime is described by an effective action, consisting of local and non-local terms \cite{Weinberg:1980gg, Starobinsky:1981ZhPmR, Barvinsky:1983vpp, Barvinsky:1985an, Barvinsky:1987uw, Barvinsky:1990up, Donoghue:1994dn}. The local action is a curvature expansion involving quantities invariant under general coordinate transformations. At second order in curvature, it is
\begin{equation}\label{eq:local_action}
    \Gamma_{L}=\int d^4x\, \sqrt{-g}\,\bigg(\frac{R}{16\pi G_N}+c_1(\mu)R^2
    +c_2(\mu)R_{\mu\nu}R^{\mu\nu}
    +c_3(\mu)R_{\mu\nu\rho\sigma}R^{\mu\nu\rho\sigma}\bigg),
\end{equation}
where $\mu$ is an energy scale. The exact values of the coefficients  $c_1,c_2,c_3$ are unknown, as they depend on the nature of the ultra-violet theory of quantum gravity. 
The non-local action is
\begin{equation}
    \Gamma_{NL}=-\int d^4 x \sqrt{-g}\bigg[\alpha R\ln\left(\frac{\Box}{\mu^2}\right)R
    +\beta R_{\mu\nu}\ln\left(\frac{\Box}{\mu^2}\right)R^{\mu\nu} + \gamma R_{\mu\nu\rho\sigma}\ln\left(\frac{\Box}{\mu^2}\right)R^{\mu\nu\rho\sigma}\bigg],
\end{equation}
where $\Box=g^{\mu\nu}\nabla_{\mu}\nabla_{\nu}$ and $\ln\left(\Box/\mu^2\right)$ is an operator with the following integral representation \cite{Donoghue:2015nba}:
\begin{equation}
    \ln\left(\frac{\Box}{\mu^2}\right)=\int_0^{+\infty}ds\, \left(\frac{1}{\mu^2+s}-\frac{1}{\Box+s}\right).
\end{equation}
The numerical values of the coefficients $\alpha,\beta,\gamma$ are calculable, see e.g. \cite{Donoghue:2014yha}. In four dimensions, the Gauss-Bonnet term
\begin{equation}
    \int d^4x\,\sqrt{-g}\left(R^2-4R_{\mu\nu}R^{\mu\nu}+R_{\mu\nu\rho\sigma}R^{\mu\nu\rho\sigma}\right)
\end{equation}
is a topological invariant. Hence, as a simplification of the equations of motion, it is possible to eliminate the Riemann tensor in the local action \eqref{eq:local_action} by redefining the coefficients as
\begin{equation}
    c_1\to\bar{c}_1=c_1-c_3, \hspace{1mm}c_2\to\bar{c}_2=c_2+4c_3, \hspace{1mm} c_3\to \bar{c}_3=0.
\end{equation}
At second order in curvature, there exists also a non-local version of the Gauss-Bonnet theorem \cite{Calmet:2018elv}:
\begin{equation}
    R_{\mu\nu\rho\sigma}\ln\left(\frac{\Box}{\mu^2}\right)R^{\mu\nu\rho\sigma}=4R_{\mu\nu}\ln\left(\frac{\Box}{\mu^2}\right)R^{\mu\nu}
    -R\ln\left(\frac{\Box}{\mu^2}\right)R.
\end{equation}
Therefore, it is possible to eliminate the Riemann tensor in the non-local action by redefining the coefficients as
\begin{equation}
\alpha\to\bar{\alpha}=\alpha-\gamma, \hspace{4mm} \beta\to\bar{\beta}=\beta+4\gamma, \hspace{4mm} \gamma\to\bar{\gamma}=0.
\end{equation}
Since we are considering an electrically charged black hole, to the purely gravitational terms we have to add the Maxwell action
\begin{equation}
    \Gamma_{M}=-\frac{1}{4}\int d^4x\,\sqrt{-g}\,F_{\mu\nu}F^{\mu\nu},
\end{equation}
where $F_{\mu\nu}=\partial_{\mu}A_{\nu}-\partial_{\nu}A_{\mu}$ is the electromagnetic tensor and $A_{\mu}$ is the electromagnetic potential. 
The full action is then
\begin{equation}\label{eq:total_action}
    \Gamma=\Gamma_{L}+\Gamma_{NL}+\Gamma_{M}=\int d^4x\,\sqrt{-g}\,\mathcal{L}.
\end{equation}
The Maxwell equations, obtained by varying \eqref{eq:total_action} with respect to $A_{\mu}$, are
\begin{equation}\label{eq:maxwell_equations}
    g^{\mu\nu}\nabla_{\mu}F_{\nu\tau}=0.
\end{equation}
In addition, there is a Bianchi identity
\begin{equation}\label{eq:bianchi}
    \nabla_{[\mu}F_{\nu\tau]}=0.
\end{equation}
The quantum corrected Einstein equations, obtained by varying \eqref{eq:total_action} with respect to the metric, are
\begin{equation}\label{eq:einstein_equations}
    \frac{1}{8\pi G_N}G_{\mu\nu}+2\left(H_{\mu\nu}+K_{\mu\nu}\right)=T_{\mu\nu}.
\end{equation}
The $T_{\mu\nu}$ on the right-hand side is the energy-momentum tensor for electromagnetism,
\begin{equation}
    T_{\mu\nu}=\frac{1}{4\pi}\left(F_{\mu\rho}{F_{\nu}}^{\rho}-\frac{1}{4}g_{\mu\nu}F_{\rho\sigma}F^{\rho\sigma}\right).
\end{equation}
On the left-hand side, $G_{\mu\nu}$ is the usual Einstein tensor
\begin{equation}
    G_{\mu\nu}=R_{\mu\nu}-\frac{1}{2}R g_{\mu\nu},
\end{equation}
while $H_{\mu\nu}$ and $K_{\mu\nu}$ are the quantum corrections to the classical Einstein's equations. The local contribution is
\begin{equation}
\begin{gathered}
    H_{\mu\nu}=\bar{c}_1\bigg(2R R_{\mu\nu}-\frac{1}{2}g_{\mu\nu}R^2-2\nabla_{\mu}\nabla_{\nu}R
    +2g_{\mu\nu}\Box R\bigg) \\+\bar{c}_2 \bigg(-\frac{1}{2}g_{\mu\nu}R_{\rho\sigma}R^{\rho\sigma}+2R^{\rho\sigma}R_{\mu\rho\nu\sigma}
    -\nabla_{\mu}\nabla_{\nu}R+\Box R_{\mu\nu}+\frac{1}{2}g_{\mu\nu}\Box R\bigg).
\end{gathered}
\end{equation}
The non-local contribution is obtained by varying $\Gamma_{NL}$. In principle one has to take the variation of the logarithm. However, $\delta/\delta g^{\mu\nu}\ln\left(\Box/\mu^2\right)$ contains terms of higher order than two in curvature \cite{Donoghue:2014yha, Codello:2015pga} so that they can be neglected here. The non-local contribution is then
\begin{equation}
\begin{gathered}
   K_{\mu\nu}=
     -2\bar{\alpha}\left(R_{\mu\nu}-\frac{1}{4}g_{\mu\nu}R+g_{\mu\nu}\Box-\nabla_{\mu}\nabla_{\nu}\right)\ln\left(\frac{\Box}{\mu^2}\right)R
    -\bar{\beta}\bigg({\delta^{\rho}}_{\mu}R_{\nu\sigma}+{\delta^{\rho}}_{\nu}R_{\mu\sigma}\\-\frac{1}{2}g_{\mu\nu}{R^{\rho}}_{\sigma}+{\delta^{\rho}}_{\mu}g_{\nu\sigma}\Box
   +g_{\mu\nu}\nabla^{\rho}\nabla_{\sigma}-{\delta^{\rho}}_{\mu}\nabla_{\sigma}\nabla_{\nu}-{\delta^{\rho}}_{\nu}\nabla_{\sigma}\nabla_{\mu}\bigg)\ln\left(\frac{\Box}{\mu^2}\right){R^{\sigma}}_{\rho}.
\end{gathered}
\end{equation}
\subsection{Solution of the equations of motion}
In this section we solve the quantum corrected Einstein and Maxwell equations using perturbation theory in the gravitational constant $G_N$. The equations are coupled together as the electromagnetic tensor $F_{\mu\nu}$ appears in the Einstein equations through the energy-momentum tensor $T_{\mu\nu}$, while the metric enters explicitly into Maxwell’s equations. 

We consider a small perturbation $g^q_{\mu\nu}$ of order $\mathcal{O}\left(G^2_N\right)$ around the classical RN solution:
\begin{equation}\label{eq:expansion}
    g_{\mu\nu}=g^{RN}_{\mu\nu}+g^{q}_{\mu\nu}.
\end{equation}
We recall that, in the canonical coordinates $x^{\mu}=\{t,r,\theta,\phi\}$, the RN solution is
\begin{equation}
\begin{gathered}
    \hspace{-2mm}ds^2_{RN}=g^{RN}_{\mu\nu}dx^{\mu}dx^{\nu}=-\left(1-\frac{2G_N M}{r}+\frac{G_N Q^2}{r^2}\right)dt^2\\
    +\left(1-\frac{2G_N M}{r}+\frac{G_N Q^2}{r^2}\right)^{-1}dr^2+
    r^2d\theta^2+r^2\sin^2\theta d\phi^2.
\end{gathered}
\end{equation}
Let us have a look at $g^q_{\mu\nu}$. We fix the gauge freedom by setting $g^q_{\theta\theta}=g^q_{\phi\phi}=0$ and define two functions $\Delta(r)$ and $\Sigma(r)$ such that 
\begin{equation}\label{eq:quantum_metric}
    ds^2_{q}=g^{q}_{\mu\nu}dx^{\mu}dx^{\nu}=-G^2_N\Delta(r)dt^2-G^2_N\Sigma(r)dr^2.
\end{equation}
The full metric is then
\begin{equation}
\begin{gathered}\label{eq:full_metric}
    ds^2=g_{\mu\nu}dx^{\mu}dx^{\nu}=
    -\left(1-\frac{2G_N M}{r}+\frac{G_N Q^2}{r^2}+G^2_N\Delta(r)\right)dt^2\\
    +\left(1-\frac{2G_N M}{r}+\frac{G_N Q^2}{r^2}+G^2_N\Sigma(r)\right)^{-1}dr^2+
    r^2d\theta^2+r^2\sin^2\theta d\phi^2.
\end{gathered}
\end{equation}
Notice that, at order $\mathcal{O}\left(G^2_N\right)$,
\begin{equation}
    \left(1-\frac{2G_N M}{r}+\frac{G_N Q^2}{r^2}\right)^{-1}-G^2_N\Sigma(r)
    =\left(1-\frac{2G_N M}{r}+\frac{G_N Q^2}{r^2}+G^2_N\Sigma(r)\right)^{-1}.
\end{equation}
The non-vanishing independent component of $F_{\mu\nu}$ for the classical RN black hole is $F^{RN}_{tr}=-F^{RN}_{rt}=Q/r^2$. Since the Maxwell and Einstein equations are coupled together, it is well possible that this component receives a correction. We thus define a function $\Omega(r)$ such that 
\begin{equation}\label{eq:expanded_F}
    F_{tr}=-F_{rt}=\frac{Q}{r^2}+G^2_N\Omega(r).
\end{equation}
To proceed, we plug the metric \eqref{eq:full_metric} and the components of the electromagnetic tensor \eqref{eq:expanded_F} into the equations \eqref{eq:maxwell_equations} - \eqref{eq:einstein_equations}, keeping only terms up to order $\mathcal{O}\left(G^2_N\right)$. Since $H_{\mu\nu}$ and $K_{\mu\nu}$ are quadratic in curvature, this implies in particular that they just need to be evaluated on the classical RN solution, i.e. \eqref{eq:einstein_equations} can be expressed schematically as
\begin{equation}\label{eq:einstein_quantum}
\frac{1}{8\pi G_N}G_{\mu\nu}\left[g^{RN}+g^{q}\right] +2 \left(H_{\mu\nu}\left[g^{RN}\right]+K_{\mu\nu}\left[g^{RN}\right]\right)=T_{\mu\nu}\left[g^{RN}+g^{q}\right].
\end{equation}
The Ricci scalar computed from the pure RN solution vanishes. Thus, in $H_{\mu\nu}$ and $K_{\mu\nu}$ only the terms containing the Ricci and Riemann tensors have to be considered. Special care is needed for the quantity ${L^{\rho}}_{\sigma}\equiv\ln\left(\frac{\Box}{\mu^2}\right){R^{\rho}}_{\sigma}$. The components of the Ricci tensor for the classical RN solution are ${R^{t}}_{t}={R^{r}}_r=-{R^{\theta}}_{\theta}=-{R^{\phi}}_{\phi}=-G_NQ^2/r^4$. Hence, it is necessary to compute $\ln\left(\Box/\mu^2\right)r^{-4}$. It was shown \cite{Calmet:2019eof} that the action of 
the logarithm on a smooth radial function $f(r)$ is
\begin{equation}
\begin{gathered}
    \ln\left(\frac{\Box}{\mu^2}\right)f(r)=\frac{1}{r}\int_0^{+\infty}dr'\,\frac{r'}{r+r'}f(r')-\lim_{\epsilon\to 0^{+}}\bigg\{ \frac{1}{r}\int_0^{r-\epsilon}dr'\,\frac{r'}{r-r'}f(r')\\
    +\frac{1}{r}\int_{r+\epsilon}^{+\infty}dr'\,\frac{r'}{r'-r}f(r')
    +2f(r)\left[\gamma_E+\ln\left(\mu\epsilon\right)\right]\bigg\},
\end{gathered}
\end{equation}
where $\gamma_E$ is the Euler-Mascheroni constant. The function $r^{-4}$ is smooth everywhere except at $r=0$, which causes some problems while evaluating the integrals at $r=0$. However, as we show in Appendix \ref{sec:appendix}, the divergences can be controlled and eventually disappear. The result is
\begin{equation}
    \ln\left(\frac{\Box}{\mu^2}\right)\frac{1}{r^4}=-\frac{2}{r^4}\left(\ln\left(\mu r\right)+\gamma_E-\frac{3}{2}\right)\equiv -\frac{2}{r^4}\ln\left(\Tilde{\mu}r\right),
\end{equation}
where we have defined for simplicity $\Tilde{\mu}=\mu\exp\left(\gamma_E-\frac{3}{2}\right)$.

We now write the explicit form of the equations. The $tt$ component of \eqref{eq:einstein_quantum} is
\begin{equation}
        -r^4\Sigma(r)-r^5\frac{d\Sigma(r)}{dr} + 64\pi Q^2\left[\left(3\bar{c}_2-5\bar{\beta}\right)+6\bar{\beta}\ln\left(\Tilde{\mu}r\right)\right]=0,
\end{equation}
which is solved by 
\begin{equation}\label{eq:solution_sigma}
\Sigma(r)=-\frac{64\pi Q^2}{r^4}\left[\bar{c}_2-\bar{\beta}+2\bar{\beta}\ln\left(\Tilde{\mu}r\right)\right].
\end{equation}
The $rr$ component of \eqref{eq:einstein_quantum} is
\begin{equation}\label{eq:rr_component}
    -r^4\Sigma(r)+64\pi Q^2\left[\bar{c}_2+2\bar{\beta}\ln\left(\Tilde{\mu}r\right)\right]-r^5\frac{d\Delta(r)}{dr}=0.
\end{equation}
Plugging \eqref{eq:solution_sigma} into \eqref{eq:rr_component} results in
\begin{equation}\label{eq:solution_delta}
    \Delta(r)=-\frac{32\pi Q^2}{r^4}\left[\bar{c}_2+2\bar{\beta}\ln\left(\Tilde{\mu}r\right)\right].
\end{equation}
The $\theta\theta$ and $\phi\phi$ components of \eqref{eq:einstein_quantum} have the form
\begin{equation}\label{eq:equation_thetaphi}
    256\pi Q^2 \bar{c}_2+128\pi Q^2\bar{\beta}\left[-1+4\ln\left(\Tilde{\mu r}\right)\right]+r^5\left[\frac{d\Delta(r)}{dr}+\frac{d\Sigma(r)}{dr}+r\frac{d^2\Delta(r)}{dr^2}\right]=0.
\end{equation}
One can check that the expressions \eqref{eq:solution_sigma}, \eqref{eq:solution_delta} for $\Sigma(r)$ and $\Delta(r)$ satisfy \eqref{eq:equation_thetaphi}.
Let us consider now the Maxwell equations. The Bianchi identity \eqref{eq:bianchi} is automatically satisfied. The $t$ component of \eqref{eq:maxwell_equations} is
\begin{equation}\label{eq:eq_maxwell}
    4r\Omega(r)+Q\left[\frac{d\Delta(r)}{dr}-\frac{d\Sigma(r)}{dr}\right]+2r^2\frac{d\Omega(r)}{dr}=0.
\end{equation}
Plugging the solutions \eqref{eq:solution_sigma} and \eqref{eq:solution_delta} into \eqref{eq:eq_maxwell} gives
\begin{equation}
\Omega(r)=\frac{16\pi Q^3 }{r^6}\left[\bar{c}_2-2\bar{\beta}+2\bar{\beta}\ln\left(\Tilde{\mu}r\right)\right].
\end{equation}
We conclude that the set of equation obtained by varying the full action \eqref{eq:total_action} has the following solution, valid at order $\mathcal{O}\left(G^2_N\right)$: 
\begin{equation}
\begin{gathered}\label{eq:final_metric}
    ds^2=-f(r)dt^2+\frac{1}{g(r)}dr^2+r^2d\theta^2+r^2\sin^2\theta d\phi^2,\\
    F_{tr}=-F_{rt}=\frac{Q}{r^2}+\frac{16\pi G^2_N Q^3}{r^6}\bigg[c_2+4c_3+2\left(\beta+4\gamma\right)\left(\ln\left(\mu r\right)+\gamma_E-\frac{5}{2}\right)\bigg]
\end{gathered}
\end{equation}
with 
\begin{equation}\label{eq:f(r)}
    f(r)=1-\frac{2G_N M}{r}+\frac{G_N Q^2}{r^2}-\frac{32\pi G^2_N Q^2}{r^4} \bigg[c_2+4c_3+2\left(\beta+4\gamma\right)\left(\ln\left(\mu r\right)+\gamma_E-\frac{3}{2}\right)\bigg],
\end{equation}
\begin{equation}\label{eq:g(r)}
   g(r)=1-\frac{2G_N M}{r}+\frac{G_N Q^2}{r^2}-\frac{64\pi G^2_N Q^2}{r^4}\Big[c_2+4c_3+2\left(\beta+4\gamma\right)\left(\ln\left(\mu r\right)+\gamma_E-2\right)\Big].
\end{equation}
In other words, both the the classical RN metric and the independent component of the electromagnetic tensor receive quantum corrections at second order in curvature. They are proportional to the charge, so in the limit $Q\to 0$ one correctly recovers the classical Schwarzschild solution. 

The components of the metric that we have just found seem to depend on the arbitrary energy scale $\mu$. However, the renormalised constants $c_1$, $c_2$, and $c_3$ carry an explicit scale dependence \cite{El-Menoufi:2015cqw}:
\begin{equation}\label{eq:RG_coefficients}
\begin{split}
    c_1(\mu)=c_1(\mu_*)-\alpha\ln\left(\frac{\mu^2}{\mu^2_*}\right),\\
    c_2(\mu)=c_2(\mu_*)-\beta\ln\left(\frac{\mu^2}{\mu^2_*}\right),\\
    c_3(\mu)=c_3(\mu_*)-\gamma\ln\left(\frac{\mu^2}{\mu^2_*}\right),\\
\end{split}
\end{equation}
where $\mu_*$ is some fixed scale where the effective theory is matched onto the full theory.
Inserting these expressions into \eqref{eq:f(r)} and \eqref{eq:g(r)}, one sees that the terms involving $\ln\mu$ cancel out. This in an indicator of the correctness of our results.
\section{Quantum gravitational corrections to the entropy of non-extremal Reissner-Nordstr\"om black holes}\label{sec:entropy}
The results obtained in the previous section are valid for a generic charge $Q$. In this section we focus on a non-extremal black hole with the additional physical condition the charge is much smaller than the mass, i.e. $Q^2 \ll GM^2$. Accordingly, in all the expressions we keep only terms up to order $\mathcal{O}\left(Q^2\right)$. 

The classical RN metric has two horizons located at the radii
\begin{equation}
\begin{split}
    &r_{+}= G_NM+\sqrt{G^2_N M^2-G_N Q^2}=2G_N M-\frac{Q^2}{2M}+\mathcal{O}\left(Q^4\right), \\ &r_{-}=G_NM-\sqrt{G^2_N M^2-G_N Q^2}=\frac{Q^2}{2M}+\mathcal{O}\left(Q^4\right).
\end{split}
\end{equation}
The entropy is generally defined in terms of $r_+$ only, namely
\begin{equation}
    S_{RN}=\frac{\pi r^2_{+}}{G_N}=\frac{A}{4G_N}-2\pi Q^2+\mathcal{O}\left(Q^4\right),
\end{equation}
where $A=16\pi G^2_N M^2$ is the classical area of the event horizon of a Schwarzschild black hole.  In this section we compute the quantum corrections to this expression.

First of all, the corrections to the metric imply a shift of the horizon radius:
\begin{equation}\label{eq:new_radius}
    r_h=2G_NM-\frac{Q^2}{2M}+\frac{8\pi Q^2}{G_NM^3}\Big[c_2+4c_3+2\left(\beta+4\gamma\right)\left(\ln\left(2G_N M\mu \right)+\gamma_E-2\right)\Big]+\mathcal{O}\left(Q^4\right).
\end{equation}
We now compute the entropy of a quantum non-extremal RN black hole using the Wald formula \eqref{eq:wald} evaluated at the new radius \eqref{eq:new_radius}. The $\epsilon_{\mu\nu}$ tensor is defined as
\begin{equation}
\epsilon_{\mu\nu}=
\begin{cases}
\sqrt{f(r)/g(r)} \hspace{10mm} \text{if} \hspace{1 mm} (\mu,\nu)=(t,r) \\ 
-\sqrt{f(r)/g(r)} \hspace{7.5 mm} \text{if} \hspace{1 mm} (\mu,\nu)=(r,t)\\
0 \hspace{30 mm} \text{otherwise}.
\end{cases}
\end{equation}
In the Wald formula, all the terms of the full effective action have to be considered, without invoking the Gauss-Bonnet theorem.  Up to orders $\mathcal{O}(Q^2)$ and $\mathcal{O}\left(G^2_N\right)$, the $rtrt$ component of the Riemann tensor and the $tt$ component of the Ricci tensor for the metric \eqref{eq:final_metric} are respectively
\begin{equation}
\begin{gathered}\label{eq:riemann}
    R^{trtr}=-\frac{2G_NM}{r^3}-\frac{3G_NQ^2}{r^4}-\frac{320\pi G^2_N Q^2}{r^6}\bigg[(c_2+4c_3)\\
    +2(\beta+4\gamma)\left(\ln(\mu r)+\gamma_E-\frac{39}{20}\right)\bigg],
\end{gathered}
\end{equation}
\begin{equation}
\begin{gathered}\label{eq:ricci}
    R^{tt}=\frac{G_NQ^2}{r^4}+\frac{2G_NMQ^2}{r^5}-\frac{192 \pi G^2_N Q^2}{r^6}\bigg[(c_2+4c_3)\\
    +2(\beta+4\gamma)\left(\ln(\mu r)+\gamma_E-\frac{11}{12}\right)\bigg]
\end{gathered}
\end{equation}
and the Ricci scalar is
\begin{equation}\label{eq:ricci_scalar}
    R=\frac{192\pi G^2_N Q^2(\beta+4\gamma)}{r^6}.
\end{equation}
Using the relations
\begin{equation}\label{eq:derivative_R}
    \frac{\partial R}{\partial R_{\mu\nu\rho\sigma}}=\frac{1}{2}\left(g^{\mu\rho}g^{\nu\sigma}-g^{\mu\sigma}g^{\nu\rho}\right),
\end{equation}
\begin{equation}\label{eq:derivative_Ricci}
    \frac{\partial\left(R_{\alpha\beta}R^{\alpha\beta}\right)}{\partial R_{\mu\nu\rho\sigma}}=\frac{1}{2}\left(g^{\mu\rho}R^{\nu\sigma}-g^{\nu\rho}R^{\mu\sigma}-g^{\mu\sigma}R^{\nu\rho}+g^{\nu\sigma}R^{\mu\rho}\right),
\end{equation}
\begin{equation}\label{eq:derivative_Riemann}
    \frac{\partial\left(R_{\alpha\beta\gamma\delta}R^{\alpha\beta\gamma\delta}\right)}{\partial R_{\mu\nu\rho\sigma}}=2R^{\mu\nu\rho\sigma}
\end{equation}
and the result of the action of $\ln\left(\Box/\mu^2\right)$ on  $1/r^3,1/r^5,1/r^6,\ln(\mu r)/r^6$, collected in Appendix \ref{sec:appendix}, we find
\begin{equation}
    S_{\text{Wald}}=-8\pi\sqrt{\frac{f(r_h)}{g(r_h)}}\int\limits_{r=r_h}d\Sigma\, \frac{\partial \mathcal{L}}{\partial R_{rtrt}}=\frac{A}{4G}-2\pi Q^2+S_0+S_2Q^2+\mathcal{O}\left(Q^4\right),
\end{equation}
where
\begin{equation}
    S_0=64\pi^2c_3+64\pi^2\gamma\Big[\ln\left(4G^2_NM^2\mu^2\right)+2\gamma_E-2\Big],
\end{equation}
\begin{multline}\label{eq:S_2}
       S_2=\frac{4\pi^2}{G_NM^2}\Big[5(c_2+4c_3)+\beta\left(10\gamma_E-21\right)+8\gamma\left(5\gamma_E-11\right)+10\left(\beta+4\gamma\right)\ln\left(2G_NM\mu\right)\Big]\\
   +\frac{64\pi^3}{9G^2_NM^4}\bigg\{54(\beta+4\gamma)\Big[c_1+2\alpha\ln\left(2G_NM\mu\right)\Big]
   +12(12\gamma_E-23)\beta\Big[c_2+2\beta\ln\left(2G_NM\mu\right)\Big]\\
   +48(48\gamma_E-97)\gamma\Big[c_3+2\gamma\ln\left(2G_NM\mu\right)\Big]
   +6\Big[c_2\gamma(96\gamma_E-185)+c_3\beta(96\gamma_E-193)\\
   +12\beta\gamma(32\gamma_E-63)\ln\left(2G_NM\mu\right)\Big]
   +36\Big[c^2_2+4c_2\beta\ln\left(2G_NM\mu\right)+4\beta^2\ln^2\left(2G_NM\mu\right)\Big]\\
   +576\Big[c^2_3+4c_3\gamma\ln\left(2G_NM\mu\right)+4\gamma^2\ln^2\left(2G_NM\mu\right)\Big]
   +288\Big[c_2c_3+2c_2\gamma\ln\left(2G_NM\mu\right)\\+2c_3\beta\ln\left(2G_NM\mu\right)
   +4\beta\gamma\ln^2\left(2G_NM\mu\right)\Big]+(\beta+4\gamma)\Big[ 9\alpha(12\gamma_E-25)
   \\+8\beta\left(3\gamma_E(6\gamma_E-23)+40+3\pi^2\right)
   +4\gamma\big(6\gamma_E\left(24\gamma_E-97\right)+331+30\pi^2\big)\Big]
\bigg\}.
\end{multline}
The first expression $S_0$ reproduces the result obtained in \cite{Calmet:2021lny} for a Schwarzschild black hole. Plugging the explicit $\mu$ dependence of the $c_i$ coefficients  \eqref{eq:RG_coefficients} into  $S_0$ and $S_2$, one can check that the terms involving $\mu$ cancel out. Hence, the corrections to the entropy are RG invariant. Our results are consistent.

We conclude this section with a remark. The second part of $S_2$ is proportional to $c_i c_j$ and $c_i \alpha_j$. In principle one could reproduce this behaviour by including in the action also terms proportional to $R^4$, like $c_iR^2c_jR^2$ and $c_iR^2\alpha_j R \ln\left(\Box/\mu^2\right)R$. However, from \eqref{eq:riemann}-\eqref{eq:ricci_scalar} the Riemann and Ricci tensor, and the Ricci scalar are proportional to $G_N$ or $G^2_N$.  This implies that the equations of motion contain terms of at least order $\mathcal{O}\left(G^3_N\right)$. For example, the equations of motion from $ \Gamma_4=\int d^4x\, R^4$ are
\begin{equation}
\begin{gathered}
4R^3 R_{\mu\nu}-\frac{1}{2}g_{\mu\nu}R^4-24R\left(\nabla_{\mu}R\right)\left(\nabla_{\nu}R\right)
-12R^2\nabla_{\mu}\nabla_{\nu}R\\+12g_{\mu\nu}R^2\Box R + 24g_{\mu\nu}R \nabla_{\rho}R\nabla^{\rho}R=0.
\end{gathered}
\end{equation}
Since we are working up to order $\mathcal{O}\left(G^2_N\right)$, we can safely neglect those terms. 
\section{Quantum gravitational corrections to temperature and pressure}\label{sec:firstlaw}
In this section we compute the quantum corrections to thermodynamic quantities of physical interest, namely temperature and pressure. The first law of thermodynamics for a Schwarzschild black hole is
\begin{equation}
    dE=dM=TdS,
\end{equation}
where $T$ and $S$ are the temperature and entropy of the black hole, respectively. The total energy of a charged black hole is $E=M+Q\Phi$, where $\Phi$ is the electric potential of the black hole. If the charge is fixed, the first law gets modified as
\begin{equation}
    dE=dM+Qd\Phi=TdS,
\end{equation}
Classically, the first non-trivial contribution to the temperature that contains the charge is a quartic term:
\begin{equation}
    T_{RN}=\frac{1}{4\pi}\Big\lvert \frac{d g^{RN}_{tt}}{dr} \Big\rvert_{r=r_{+}}=\frac{1}{8\pi G_N M}-\frac{Q^4}{128\pi G^3_NM^5}+\mathcal{O}\left(Q^6\right).
\end{equation}
The corrections to the metric imply a correction to the temperature of order $\mathcal{O}\left(Q^2\right)$:
\begin{equation}
\begin{gathered}
    T=\frac{1}{4\pi}\sqrt{\frac{df(r)}{dr}\frac{dg(r)}{dr}}\bigg\rvert_{r=r_h}=\frac{1}{8\pi G_N M}
    +\frac{Q^2}{4G^3_NM^5}\Big[2(c_2+4c_3)\\+\beta\left(4\gamma_E-9\right)+4\gamma\left(4\gamma_E-9\right)
    +4\left(\beta+4\gamma\right)\ln\left(2G_NM\mu\right)\Big]+\mathcal{O}\left(Q^4\right).
\end{gathered}
\end{equation}
The electric potential is
\begin{equation}
\Phi=\int_{r_h}^{+\infty} dr'\, F_{tr}=\int_{r_h}^{+\infty} dr'\, \left(\frac{Q}{{r'}^2}+G^2_N\Omega(r')\right)=\frac{Q}{2G_NM}+\mathcal{O}\left(Q^3\right),
\end{equation}
meaning that it does not receive corrections at order $\mathcal{O}\left(Q^2\right)$.

Even at the classical level, one finds that $TdS/dM-Qd\Phi/dM=1+\Delta(M,Q)$ i.e. the unity plus some additional terms. In order to preserve the first law, one assumes that the charged black hole has a pressure $P$ such that
\begin{equation}
    T\frac{dS}{dM}-Q\frac{d\Phi}{dM}-P\frac{dV}{dM}=1+\Delta(M,Q),
\end{equation}
where $V=4/3\pi r^3_h$ is the volume of the black hole.
The pressure is then given by
\begin{equation}\label{eq:pressure}
    P=-\frac{\Delta(M,Q)}{\frac{dV}{dM}}=-\frac{T\frac{dS}{dM}-Q\frac{d\Phi}{dM}-1}{\frac{dV}{dM}}.
\end{equation}
The pressure on the outer horizon of a classical RN black hole is negative and proportional to the charge squared \cite{Wei:2009zzc}:
\begin{equation}
    P_{RN}\sim - \frac{Q^2}{r^4_{+}}.
\end{equation}
Evidently, a Schwarzschild black hole does not have a pressure at the classical level. However, quantum effects do create a pressure \cite{Calmet:2021lny}.
Using \eqref{eq:pressure} we find that the pressure of a non-extremal RN black hole, including the quantum corrections, is 
\begin{equation}
    P=-\frac{Q^2}{64\pi G^4_NM^4}-\frac{\gamma}{2G^4_N M^4}+P_2Q^2+\mathcal{O}\left(Q^4\right)
\end{equation}
with
\begin{multline}
   P_2=\frac{1}{32G^5_NM^6}\Big[c_2+4c_3+2\beta(\gamma_E-4)+8\gamma(\gamma_E-5)+2(\beta+4\gamma)\ln(2G_NM\mu)\Big]
    \\+\frac{\pi}{9G^6_NM^8}\bigg\{54(\beta+4\gamma)\Big[c_1+2\alpha\ln\left(2G_NM\mu\right)\Big]
    +24(6\gamma_E-13)\beta\Big[c_2+2\beta\ln\left(2G_NM\mu\right)\Big]\\
    +768(3\gamma_E-7)\gamma\Big[c_3+2\gamma\ln\left(2G_NM\mu\right)\Big]
    +6\Big[c_2\gamma(96\gamma_E-215)+c_3\beta(96\gamma_E-217)\\
    +96\beta\gamma(4\gamma_E-9)\ln\left(2G_NM\mu\right)\Big]
     +36\Big[c^2_2+4c_2\beta\ln\left(2G_NM\mu\right)+4\beta^2\ln^2\left(2G_NM\mu\right)\Big]\\
   +576\Big[c^2_3+4c_3\gamma\ln\left(2G_NM\mu\right)+4\gamma^2\ln^2\left(2G_NM\mu\right)\Big]
   +288\Big[c_2c_3+2c_2\gamma\ln\left(2G_NM\mu\right)\\+2c_3\beta\ln\left(2G_NM\mu\right)
    +4\beta\gamma\ln^2\left(2G_NM\mu\right)\Big]+(\beta+4\gamma)\Big[36\alpha(3\gamma_E-7)\\ +2\beta\big(8\gamma_E(9\gamma_E-39)+229+12\pi^2\big)
    +\gamma\big(192\gamma_E(3\gamma_E-14)+2095+120\pi^2\big)\Big]  
    \bigg\}.
    \end{multline}
The term $-\gamma/2G^4_NM^4$ is the only quantum correction for a Schwarzschild black hole. A charged black hole has additional quantum corrections proportional to the electric charge.
\section{Conclusions and outlook}
In this paper we have studied the quantum gravitational corrections to the Wald entropy of a four dimensional, non-extremal Reissner-Nordstr\"om black hole, starting from an effective action for quantum gravity supplemented by the Maxwell action. The corrections to the entropy are RG invariant.  Furthermore, contrary to what happens for the Schwarzschild metric, we have shown that the classical RN solution receives quantum corrections at second order in curvature. These corrections shift the position of the event horizon. Finally, we computed the quantum gravitational corrections to the temperature and pressure. All these corrections are very small for astrophysical black holes, for which $Q^2\ll GM^2$. 

The research presented in this paper naturally lends itself to further generalisations. For instance, one could try to compute the quantum gravitational corrections to the entropy of a charged rotating (Kerr-Newman) black hole, which is the most realistic model of an astrophysical black hole.  Alternatively, one could also have a look at AdS black holes, whose importance lies in the context of the AdS/CFT correspondence. 
\section{Acknowledgements}
The author thanks the Bethe Center for Theoretical Physics (BCTP) of the University of
Bonn for its hospitality, Saurabh Natu for valuable advice and discussions about this research and the anonymous reviewers for their comments, which improved the final version of the manuscript. This work was supported by the Bonn-Cologne Graduate School of Physics and Astronomy (BCGS).
\newpage
\appendix
\section{Action of $\ln\left(\frac{\Box}{\mu^2}\right)$ on radial functions}\label{sec:appendix}
In this appendix we collect the result of the action of $\ln\left(\Box/\mu^2\right)$ on various radial functions $f(r)$. If $r>0$, $f(r)\neq0$ and $\exists \epsilon >0$ such that $f(r')$ is smooth for $\lvert r-r' \rvert \leq \epsilon$, then  \cite{Calmet:2019eof} 
\begin{equation}\label{eq:integral_appendix}
\begin{gathered}
    \ln\left(\frac{\Box}{\mu^2}\right)f(r)=\frac{1}{r}\int_0^{+\infty}dr'\,\frac{r'}{r+r'}f(r')-\lim_{\epsilon\to 0^{+}}\bigg\{ \frac{1}{r}\int_0^{r-\epsilon}dr'\,\frac{r'}{r-r'}f(r')\\
    +\frac{1}{r}\int_{r+\epsilon}^{+\infty}dr'\,\frac{r'}{r'-r}f(r')
    +2f(r)\left[\gamma_E+\ln\left(\mu\epsilon\right)\right]\bigg\}.
\end{gathered}
\end{equation}
In order to derive the equations of motion from the effective action \eqref{eq:total_action} in Section \ref{sec:equations_metric}, the non-local contribution $K_{\mu\nu}$ has to be evaluated on the classical RN solution. Since the components of the Ricci tensor for the classical RN solution are ${R^t}_t={R^r}_{r}=-{R^\theta}_{\theta}=-{R^\phi}_{\phi}=-G_NQ^2/r^4$, the following quantity has to be computed:
\begin{equation}
    \ln\left(\frac{\Box}{\mu^2}\right)\frac{1}{r^4}.
\end{equation}
Owing to the singularity at $r=0$, the integrals in \eqref{eq:integral_appendix} need to be regularised. A way to achieve this is by shifting the denominators of the problematic integrands via $r'\to r'\pm i\epsilon$ and then taking the real part of the result, as $\ln(\Box/\mu^2)f(r)\in \mathbb{R}$. Thus
\begin{equation}
\begin{gathered}
     \ln\left(\frac{\Box}{\mu^2}\right)\frac{1}{r^4}=\text{Re}\lim_{\epsilon\to 0^+}\bigg\{\frac{1}{r}\int_0^{+\infty}dr'\,\frac{r'}{r+r'}\frac{1}{\left(r'-i\epsilon\right)^4}- \frac{1}{r}\int_0^{r-\epsilon}dr'\,\frac{r'}{r-r'}\frac{1}{\left(r'+i\epsilon\right)^4}\\
    -\frac{1}{r}\int_{r+\epsilon}^{+\infty}dr'\,\frac{r'}{r'-r}\frac{1}{{r'}^4}
    -\frac{2}{r^4}\left[\gamma_E+\ln\left(\mu\epsilon\right)\right]\bigg\}.
\end{gathered}
\end{equation}
The divergences cancel out:
\begin{equation}
\begin{gathered}
  \ln\left(\frac{\Box}{\mu^2}\right)\frac{1}{r^4}=\text{Re}\Big[\frac{1}{r}\left(\frac{3}{r^3}+i\frac{\pi}{r^3}-\frac{2\ln r}{r^3}+\frac{2\ln\epsilon}{r^3}\right)-\frac{2\gamma_E}{r^4}-\frac{2\ln\epsilon}{r^4}-\frac{2\ln\mu}{r^4}\Big]=\\
  =-\frac{2}{r^4}\left(\ln\left(\mu r\right)+\gamma_E-\frac{3}{2}\right).
\end{gathered}
\end{equation}
In Section \ref{sec:entropy} we computed the corrections to the entropy using the Wald formula. The derivative of the effective action with respect to the Riemann tensor contains contributions from the Riemann and Ricci tensors of the metric with quantum corrections \eqref{eq:f(r)}, \eqref{eq:g(r)}. These tensors contain the radial functions $r^{-3},r^{-5},r^{-6}, \ln(\mu r)r^{-6}$. Hence, it is necessary to know the action of $\ln\left(\Box/\mu^2\right)$ on them. 
Let us consider for example $\ln\left(\Box/\mu^2\right)r^{-3}$. We regularise the integrals in \eqref{eq:integral_appendix} as
\begin{equation}
\begin{gathered}
     \ln\left(\frac{\Box}{\mu^2}\right)\frac{1}{r^3}=\lim_{\epsilon\to 0^+}\bigg\{\frac{1}{r}\int_{\sqrt{\epsilon r}}^{+\infty}dr'\,\frac{r'}{r+r'}\frac{1}{{r'}^3}- \frac{1}{r}\int_{\sqrt{\epsilon r}}^{r-\sqrt{\epsilon r}}dr'\,\frac{r'}{r-r'}\frac{1}{{r'}^3}
    \\-\frac{1}{r}\int_{r+\sqrt{\epsilon r}}^{+\infty}dr'\,\frac{r'}{r'-r}\frac{1}{{r'}^3}
    -\frac{2}{r^3}\left[\gamma_E+\ln\left(\mu\epsilon\right)\right]\bigg\}.
\end{gathered}
\end{equation}
Again, all divergences cancel out and the result is
\begin{equation}
\ln\left(\frac{\Box}{\mu^2}\right)\frac{1}{r^3}=-\frac{2}{r^3}\left(\ln\left(\mu r\right)+\gamma_E-1\right).
\end{equation}
With analogous calculations we find
\begin{equation}
    \ln\left(\frac{\Box}{\mu^2}\right)\frac{1}{r^5}=-\frac{2}{r^5}\left(\ln(\mu r)+\gamma_E-\frac{47}{12}\right),
\end{equation}
\begin{equation}
    \ln\left(\frac{\Box}{\mu^2}\right)\frac{1}{r^6}=-\frac{2}{r^6}\left(\ln(\mu r)+\gamma_E-\frac{25}{12}\right),
\end{equation}
\begin{equation}
     \ln\left(\frac{\Box}{\mu^2}\right)\frac{\ln\left(\Tilde{\mu}r\right)}{r^6}=\\=-\frac{2\ln\left(\tilde{\mu}r\right)}{r^6}\left(\ln(\mu r)+\gamma_E-\frac{25}{12}\right)+\\+\frac{1}{r^6}\left(\frac{205}{72}-\frac{\pi^2}{3}\right).
\end{equation}

\bibliographystyle{utphys}
\bibliography{references}

\end{document}